\documentclass{article}
\usepackage{ijcai25}

\usepackage{times}
\usepackage{soul}
\usepackage{url}
\usepackage{placeins}
\usepackage{float}
\usepackage{color}
\usepackage[hidelinks]{hyperref}
\usepackage[utf8]{inputenc}
\usepackage[small]{caption}
\usepackage{graphicx}
\usepackage{amsmath}
\usepackage{amsthm}
\usepackage{booktabs}
\usepackage{algorithm}
\usepackage{algorithmic}
\usepackage[switch]{lineno}
\usepackage{array}
\usepackage{makecell}

\title{Solving Copyright Infringement on Short Video Platforms: Novel Datasets and an Audio Restoration Deep Learning Pipeline}

\author{
Minwoo Oh$^1$
\and
Minsu Park$^{2,}$\footnotemark[1]
\and
Eunil Park$^{1,2,3,}$\thanks{Corresponding authors}\\
\affiliations
$^1$Department of MetaBioHealth, Sungkyunkwan University, Korea\\
$^2$Department of Applied Artificial Intelligence, Sungkyunkwan University, Korea\\
$^3$Department of Computer Science and Engineering, Jaume I University, Spain\\
\emails
\{mwoh0208, mspark501\}@g.skku.edu,
eunilpark@skku.edu
}

\begin{document}

\maketitle

\begin{abstract}
    Short video platforms like YouTube Shorts and TikTok face significant copyright compliance challenges, as infringers frequently embed arbitrary background music (BGM) to obscure original soundtracks (OST) and evade content originality detection. To tackle this issue, we propose a novel pipeline that integrates Music Source Separation (MSS) and cross-modal video-music retrieval (CMVMR). Our approach effectively separates arbitrary BGM from the original OST, enabling the restoration of authentic video audio tracks. To support this work, we introduce two domain-specific datasets: \textbf{OASD-20K} for audio separation and \textbf{OSVAR-160} for pipeline evaluation. OASD-20K contains 20,000 audio clips featuring mixed BGM and OST pairs, while OSVAR-160 is a unique benchmark dataset comprising 1,121 video and mixed-audio pairs, specifically designed for short video restoration tasks. Experimental results demonstrate that our pipeline not only removes arbitrary BGM with high accuracy but also restores OSTs, ensuring content integrity. This approach provides an ethical and scalable solution to copyright challenges in user-generated content on short video platforms.
\end{abstract}

\section{Introduction}

\begin{figure}[h]
    \centering
    \includegraphics[width=\columnwidth]{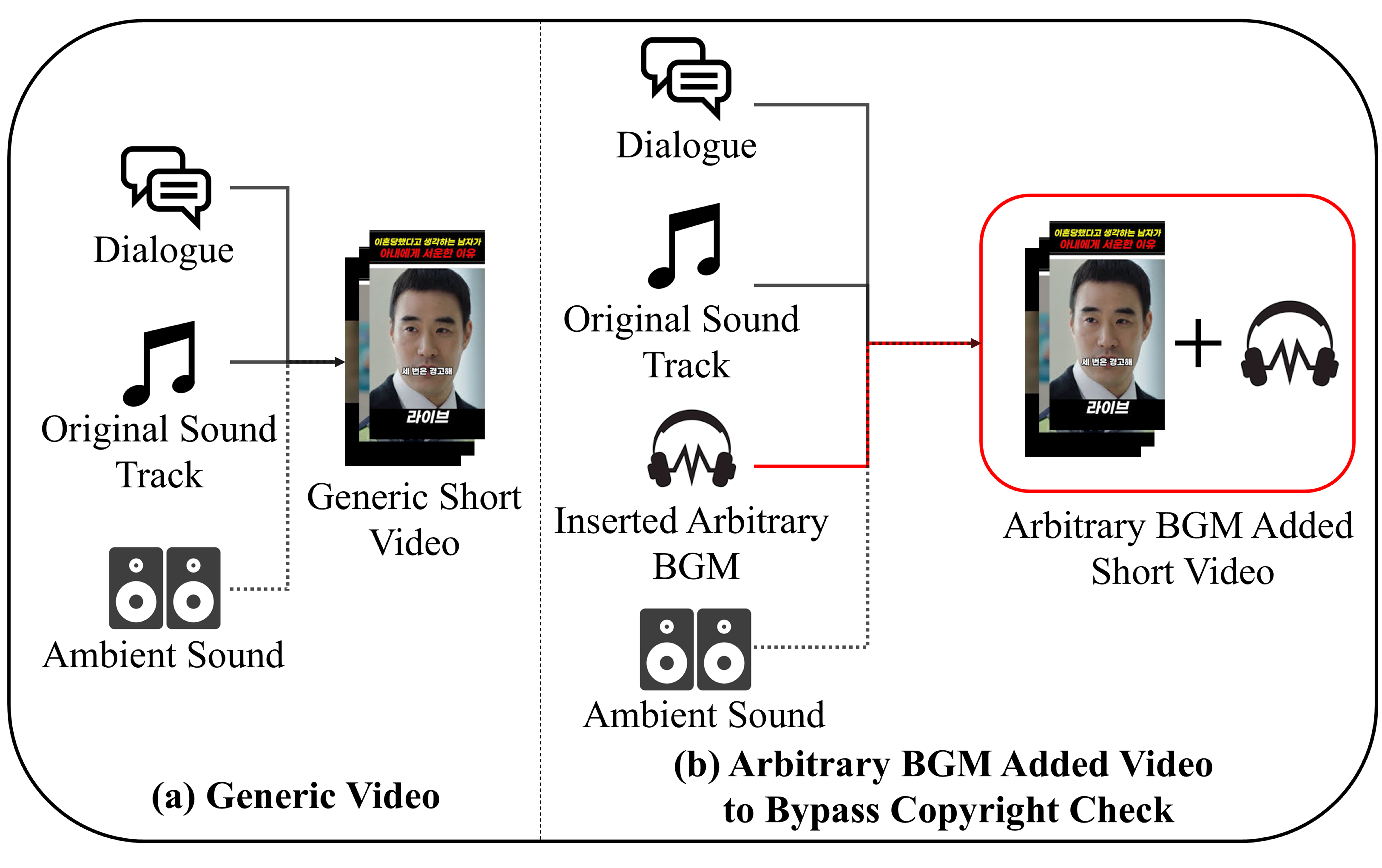} 
    \caption{Comparison of generic and arbitrarily BGM-added videos}
    \label{fig:problem}
    \vspace{-9pt}
\end{figure}

Short video platforms like YouTube Shorts, TikTok, and Instagram Reels have transformed content consumption, driving global growth and engagement. In 2023, the market for these platforms was valued at \$26.69 billion, with projections reaching \$289.52 billion by 2032~\cite{marketresearchfuture:2023}. These platforms captivate users, with 66\% expressing a preference for short-form videos and spending over an hour daily engaging with them~\cite{demandsage:2023}.

Despite their popularity, user-generated content introduces significant copyright challenges. As presented in Figure~\ref{fig:problem}, users frequently overlay arbitrary background music (BGM) on the generic short videos, disrupting the original audio structure and complicating copyright compliance~\cite{audiowatermark,hasanovic2022can,youtubehelp:shorts}. This practice masks critical audio elements, such as dialogue or original soundtracks (OST), and undermines the accuracy of audio-based near-duplicate video retrieval systems~\cite{avgoustinakis2021audio}, which are essential for identifying infringements. Addressing these issues demands advanced methods capable of removing arbitrary BGM effectively while preserving the integrity of the original audio.

Music Source Separation (MSS)~\cite{cano2018musical}, such as Spleeter~\cite{spleeter} Demucs~\cite{demucs}, presented impressive performance in isolating overlapping audio sources. In our pipeline, MSS is employed for initial audio separation. To tackle the complex task of OST-BGM separation, we further incorporate the state-of-the-art BS-RoFormer model~\cite{lu2024music}, which leverages a Band-Split module to process audio across multiple frequency bands and Rotary Position Embedding (RoPE) to capture temporal and frequency dependencies. This makes BS-RoFormer particularly well-suited for highly blended OST-BGM scenarios. However, while MSS excels at separating mixed sources, it cannot distinguish the OST from BGM, necessitating additional methods for accurate audio restoration.

To address this limitation, we integrate Cross-Modal Video-Music Retrieval (CMVMR) technologies~\cite{jin2021cross,pretet2021cross,won2021multimodal,yi2021cross} into our pipeline to align restored OSTs with their corresponding video content. Specifically, we fine-tune the UT-CMVMR model~\cite{unified}, which uses a Video Conformer and Music Conformer enhanced with Optical Flow and Rhythm Quantization to extract spatial-temporal and rhythmic features. Using cross-attention mechanisms and optimizing with Triplet Loss, the model enhances the similarity between video and music embeddings without relying on metadata or tags, making it ideal for short-video environments.

To overcome the scarcity of datasets tailored for arbitrary BGM removal in short video platforms, we introduce two novel datasets: the \textbf{Oh-Audio Separation Dataset 20K} (\textbf{OASD-20K}) and \textbf{Oh-Short Video Audio Restoration Dataset 160} (\textbf{OSVAR-160}). OASD-20K, designed for mixed-music separation, comprises 20,000 four-second audio clips, which are formed by combining OST tracks from the Korean TV series with BGM tracks from the YouTube Audio Library. This dataset reflects real-world OST-BGM blending complexities and provides high-quality training data for MSS research. OSVAR-160 consists of 160 one-minute video clips sourced from high-quality Korean TV series paired with distinct BGM tracks. It serves as a benchmark for evaluating pipeline performance in removing arbitrary BGM and restoring OST integrity.

This study introduces a novel and optimized pipeline for short video platforms like YouTube Shorts by integrating advanced audio separation and cross-modal retrieval techniques. Our approach effectively removes arbitrary BGM and restores OST integrity, addressing critical copyright challenges. The pipeline is rigorously trained and evaluated using our domain-specific datasets, OASD-20K and OSVAR-160, setting a new benchmark for arbitrary BGM removal in short video restoration tasks. The contributions are as follows:
\begin{itemize}
\item \textbf{A novel pipeline}: We introduce an innovative approach that integrates advanced MSS and CMVMR techniques to effectively remove arbitrary BGM, addressing the unique challenges faced by short video platforms.
\item \textbf{Two comprehensive datasets}: We present \textbf{OASD-20K}, a dataset designed for training and evaluating mixed-music separation, and \textbf{OSVAR-160}, a benchmark dataset for assessing arbitrary BGM removal from video content. These datasets bridge critical gaps in audio separation research for short video applications.
\item \textbf{Real-world applicability}: Through rigorous experimentation, we demonstrate the pipeline's ability to preserve audio integrity, enhance content quality, and address copyright compliance on platforms such as YouTube Shorts and TikTok.
\end{itemize}

By advancing techniques in OST restoration, mixed-music separation, and audio-based video retrieval, this research addresses pressing challenges in copyright compliance and content integrity. The proposed solutions are scalable, supporting sustainable growth and fostering ethical content practices in user-generated environments~\cite{zhao2024liability}.

\section{Related Work}
\subsection{Copyright Challenges in Short Videos}
The rapid growth of platforms like YouTube and TikTok presents concerns about copyright infringement~\cite{al2024regulating,blanding2024navigating}, spurring the development of advanced detection techniques to address unauthorized content modifications.

Many visual-based duplication detection methods utilize temporal consistency and feature-based analysis to identify duplicate videos~\cite{henry2024fast,han2024sfd,udaypur2024novel}. While these approaches are effective at identifying visual similarities, they often falter when faced with audio modifications—a prevalent tactic employed to bypass detection systems. To address this gap, audio-based retrieval methods leverage audio similarity learning techniques to retrieve near-duplicate videos~\cite{avgoustinakis2021audio,liu2010coherent,saracoglu2009content}.

Short video platforms introduce unique challenges, including inconsistent metadata and highly dynamic user-generated content. Addressing these complexities requires advanced audio processing methods, such as MSS, to isolate and remove arbitrary BGM, thereby restoring the original audio.

\subsection{Music Source Separation Approaches}
MSS aims to isolate individual components, such as vocals and background music, from mixed audio signals. Traditional signal processing-based methods rely on harmonic structures and time-frequency representations~\cite{cohen2001speech,virtanen2007monaural}, but they often struggle with complex audio mixtures.

Recent advancements in deep learning revolutionized MSS. Waveform-based approaches, such as Demucs~\cite{demucs}, operate directly on raw audio signals, capturing fine-grained temporal details to achieve high-quality separation. Spectrogram-based methods~\cite{huang2014deep,jansson2017singing}, on the other hand, analyze time-frequency representations and utilize CNNs and RNNs for precise audio component isolation~\cite{cho2014learning,sharif2014cnn}. The introduction of band-split techniques and positional embeddings~\cite{lu2024music} further enhanced separation accuracy, particularly in scenarios involving overlapping audio sources.

These advancements address the challenges of arbitrary BGM separation in short-form video content, enabling accurate separation of OST from BGM.

\subsection{Video-Music Retrieval Approaches}
Other early methods, such as the Coherent Bag-of-Audio-Words (CBoAW)~\cite{liu2010coherent}, employed audio fingerprinting techniques to detect near-duplicate content and establish relationships between videos and background music. While effective in specific scenarios, these techniques were heavily dependent on handcrafted features, making them inflexible in handling diverse content genres or dynamic environments. As these limitations became evident, researchers began exploring more sophisticated approaches.

Video-Music Retrieval (VMR) methodologies introduced content-based approaches to align audio with video content, leveraging audiovisual features like rhythm and tempo to identify patterns. For instance, CBVMR~\cite{hong2018cbvmr} demonstrated robust performance in metadata-scarce environments by directly processing video frames and audio signals. However, these methods often incur substantial costs, limiting their scalability in real-world applications.

To address these challenges, cross-modal retrieval methods emerged as a powerful alternative. Unlike traditional approaches, which process audiovisual features independently and demand significant computational resources, cross-modal methods embed video and audio data into a shared latent space. This reduces the complexity of feature extraction and alignment while maintaining high performance~\cite{gu2023dual,li2019query}. For example, UT-CMVMR~\cite{unified} employs Optical Flow to capture temporal video dynamics and Rhythm Quantization to analyze audio patterns, enabling efficient video-music alignment even in dynamic environments. These methods are well-suited for short video platforms where metadata is sparse.

These advancements in video-music synchronization address the unique demands of short-form video platforms by bypassing metadata reliance and improving scalability for real-world applications.

\section{Datasets}
To comprehensively train, fine-tune, and evaluate ours, we established two novel datasets and adapted one off-the-shelf dataset, each tailored to a specific module of the pipeline. The \textbf{OASD-20K} dataset was developed to train and fine-tune mixed-music separation models, addressing key challenges in MSS research. For the video-music alignment module, we adapted the HIMV-200K dataset to create a CMVMR fine-tuning dataset, enabling precise training of cross-modal video-music retrieval models. Finally, the \textbf{OSVAR-160} dataset was designed as a benchmark for evaluating the performance of the entire pipeline under real-world conditions, ensuring its applicability to short video scenarios.

\subsection{Mixed-Music Separator Fine-tuning Dataset (OASD-20K)}
\textbf{OASD-20K} is the first dataset specifically designed for mixed-music separation tailored to short video platforms. It addresses real-world scenarios where OST and BGM are blended, providing a robust benchmark for advancing MSS research. The dataset emphasizes ethical content usage and copyright compliance, aligning with the needs of modern content platforms. The construction process involves mixing OST and BGM tracks to generate diverse, realistic audio data representative of real-world use cases.

\subsubsection{Audio Sources and Preparation}
The OST tracks were randomly curated from a Korean TV series OST playlist\footnote{\url{https://www.youtube.com/playlist?list=RDCLAK5uy_miHnhrZJWX7HvNEX7CaSZzfk77iwWhBLI}}, while the BGM tracks were sourced from the YouTube Audio Library\footnote{\url{https://www.youtube.com/audiolibrary}}—a diverse and copyright-compliant repository commonly exploited by copyright infringers to obscure original OSTs. A total of 50 tracks were sampled from each source, with each track segmented into up to 10 four-second clips. This process yielded 500 unique OST clips and 500 unique BGM clips, creating a rich and varied dataset tailored for mixed-music separation tasks.

\subsubsection{Mixing Process and Dataset Splits}
The OST and BGM tracks were randomly mixed to generate 20,000 four-second audio clips, resulting in a dataset with approximately 22.2 hours of audio content. The dataset was split into training (16,000 clips; about 17.8 hours), validation (2,000 clips; about 2.2 hours), and test (2,000 clips; about 2.2 hours) sets, adhering to an 8:1:1 ratio. To ensure consistent volume levels across all clips, normalization was applied to -23 LUFS, following standard audio preprocessing practices\cite{ebu2011loudness}.

\begin{figure*}[!t] 
    \centering
    \includegraphics[width=.99\textwidth]{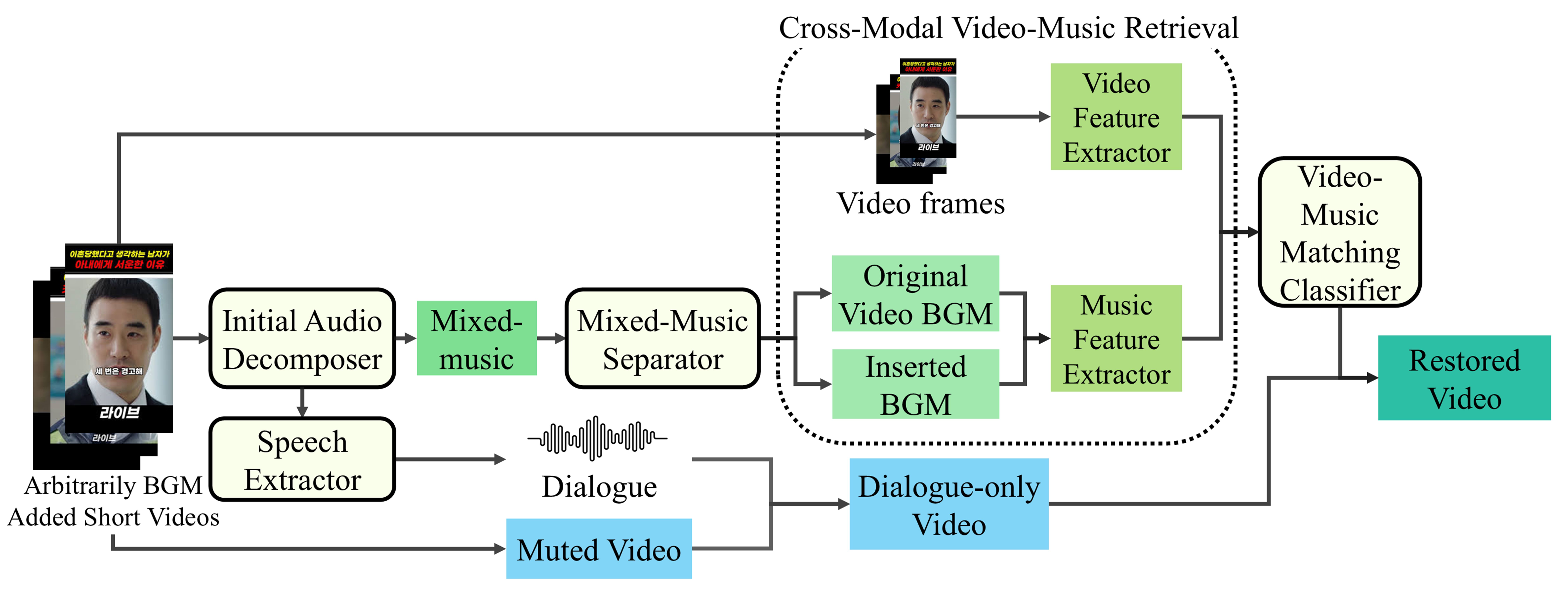} 
    \caption{Overview of the complete pipeline consisting of four key modules: Background Music Separator, Speech Extractor, Mixed-Music Separator, and Video-Music Matching Classifier. The pipeline is designed to 1) remove arbitrarily inserted BGM, 2) restore the original OST, and 3) classify appropriate background music for video content.}
    \label{fig:end_to_end_pipeline}
    \vspace{-6pt}
\end{figure*}

\subsection{Cross-modal Video-Music Retrieval Fine-tuning Dataset}
The CMVMR fine-tuning dataset was derived from the HIMV-200K dataset~\cite{hong2018cbvmr}, which contains 205,000 video-music pairs designed to ensure strong contextual alignment between video content and corresponding music. For this study, a subset of 1,000 video-music pairs was randomly selected to align with the pipeline's specific objectives. This refined subset enables the fine-tuning of UT-CMVMR, allowing for the classification of music that aligns seamlessly with video content. By eliminating tag-based dependencies, this approach reduces bias and enhances the model’s adaptability to metadata-scarce environments.

\subsubsection{Preprocessing}
The preprocessing pipeline follows the approach utilized in UT-CMVMR~\cite{unified}, ensuring compatibility with its architecture and maintaining input consistency. Each of the 1,000 video-music pairs was segmented into 28-second units, discarding shorter segments to preserve uniformity. These 28-second units were further divided into 4-second sub-clips, increasing the dataset size to support robust model training and evaluation. Video frames were resized to 224×224 pixels, sampled at 1 frame per second (FPS), and concatenated into 896×224 matrices. Corresponding audio clips were transformed into 398×80 Mel-spectrograms, ensuring seamless integration with the model's input requirements. The dataset was divided into training (9,044 pairs), validation (1,130 pairs), and test (1,131 pairs) sets, maintaining an 8:1:1 split ratio.

\subsection{Pipeline Performance Evaluation Dataset}
\textbf{OSVAR-160} is a benchmark dataset to evaluate the pipeline's effectiveness in removing arbitrary BGM from short video content. Unlike \textbf{OASD-20K} and HIMV-200K, which focus on fine-tuning specific pipeline modules, it is designed to assess the pipeline's end-to-end performance in real-world scenarios. Arbitrary BGM tracks are overlaid on video content to simulate realistic short video environments, providing a challenging testbed for evaluating the system's ability to restore original audio integrity.

\subsubsection{Video and Audio Sources}
The video data was curated from publicly available clips of Korean TV series uploaded by tvN\footnote{\url{https://www.youtube.com/@tvNDRAMA_official}}, a premier South Korean broadcasting channel renowned for its high-quality productions, compelling storytelling, and superior production value. A total of 160 one-minute clips (approximately 2.67 hours) were selected from their YouTube uploads. For the background music (BGM), distinct tracks were sourced from the YouTube Audio Library, ensuring no overlap with OASD-20K to maintain data independence. To ensure uninterrupted coverage, only tracks longer than one minute were chosen, guaranteeing that the BGM fully spans the video duration. This approach ensures consistent and uninterrupted audio content, facilitating robust evaluation of BGM removal tasks.

\subsubsection{Dataset Composition}
The video-audio pairs, with overlaid BGM, were segmented into 4-second clips, producing a total of 1,121 video-audio pairs. Audio normalization was applied following the same procedure as OASD-20K (-23 LUFS) to maintain consistent volume levels across the dataset. This segmentation process created a diverse set of short clips that include both OST and arbitrary BGM, simulating the complex challenges of short video platforms. The OSVAR-160 benchmark dataset is designed to highlight these challenges, providing a comprehensive testbed to evaluate the effectiveness of the proposed pipeline in arbitrary BGM removal and OST restoration.

\section{Method}
Our proposed method is designed to address the challenges of removing arbitrarily inserted BGM in short video content. The pipeline comprises four interconnected modules: the Background Music Separator, Speech Extractor, Mixed-Music Separator, and Video-Music Matching Classifier. An overview of the pipeline is depicted in Figure~\ref{fig:end_to_end_pipeline}.

\subsection{Initial Audio Separation: Background Music and Speech Extraction}
The first stage of the pipeline focuses on separating mixed audio (arbitrary BGM and original OST) and speech components from the video’s audio track, providing clean inputs for downstream tasks. This stage consists of two key modules: 

\paragraph{(1) Background Music Separator:} The Background Music Separator utilizes the U-Net–based Demucs model~\cite{demucs} to decompose audio into its constituent tracks, such as vocals, bass, and drums. Demucs operates in the time domain, leveraging recurrent layers to capture long-term dependencies~\cite{ronneberger2015u}, which ensures high-quality separation. The separated BGM components are then consolidated into a single ``Mixed-Music'' track using Pydub, creating a unified input for subsequent modules.

\begin{figure*}[!t] 
    \centering
    \includegraphics[width=.95\textwidth]{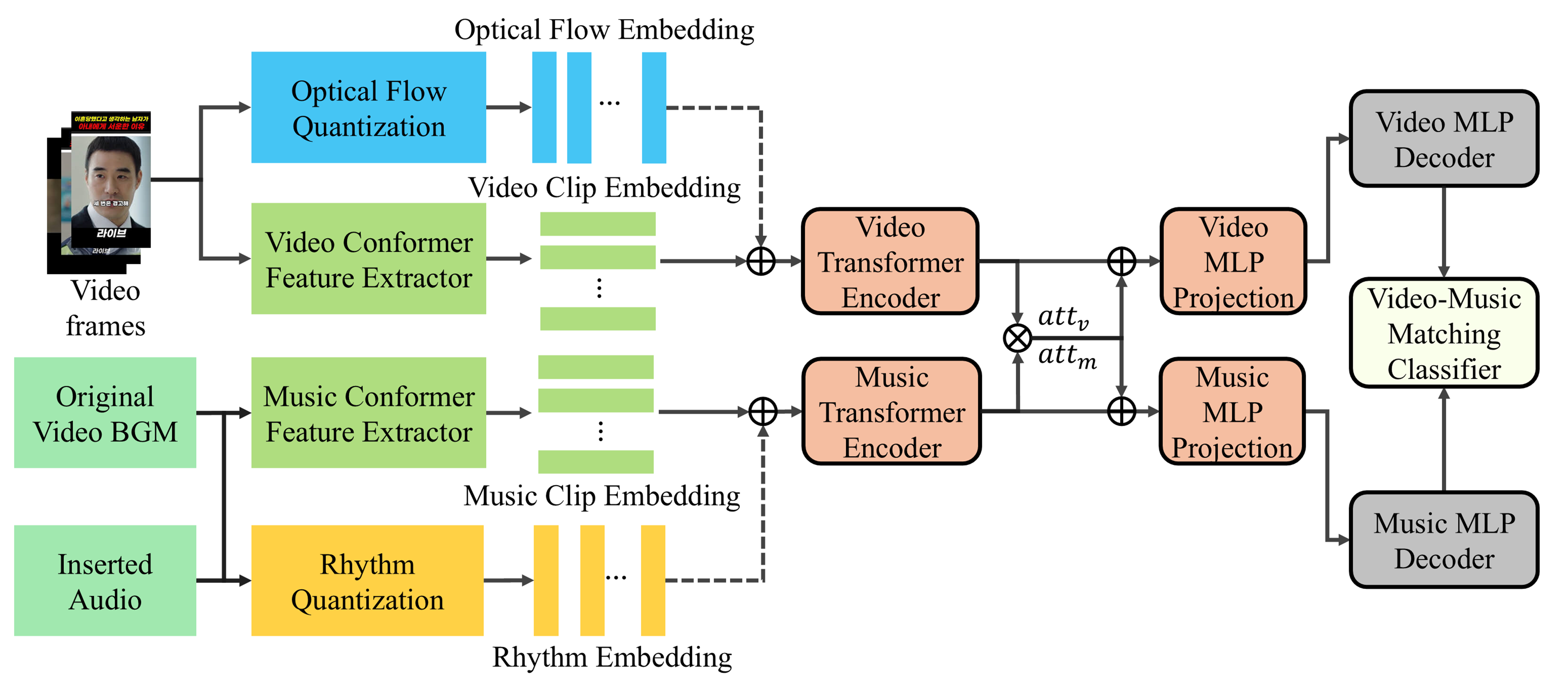} 
    \caption{Detailed architecture of the cross-modal video-music retrieval module within the pipeline.}
    \label{fig:cross_modal_pipeline}
       \vspace{-6pt}
\end{figure*}

\paragraph{(2) Speech Extractor:} The Speech Extractor applies the Pyannote Audio VAD~\cite{bredin2020pyannote} to the vocal output generated by Demucs, accurately detecting and isolating speech. By leveraging recurrent neural networks (RNNs) and temporal convolutional networks (TCNs)~\cite{soo2017interpretable}, the module processes overlapping audio frames with contextual embeddings, ensuring robust extraction of dialogue even in noisy or complex audio environments. This precise speech isolation is crucial for maintaining the integrity of spoken content during BGM removal.

This two-step process ensures precise separation of BGM and dialogue. While Demucs isolates the overall vocal track, the VAD module further refines the output by removing residual noise and non-speech elements, yielding clear and isolated dialogue. The refined speech tracks are then synchronized with the muted video, resulting in a ``Dialogue-only Video'' that provides a clean reference for subsequent audio restoration tasks. This initial stage produces clean, well-separated audio inputs, forming a solid foundation for the downstream tasks of mixed-music separation and video-music matching.

\subsection{Mixed-Music Separator}
The BS-RoFormer model~\cite{lu2024music} was adapted to separate OST and BGM in short video platforms, utilizing the OASD-20K dataset to align with real-world OST-BGM blending scenarios. Unlike previous models focused on multi-component separation (e.g., vocals, bass), BS-RoFormer’s Band-Split module processes audio signals across parallel frequency bands~\cite{luo2023music}, ensuring high-resolution spectral details even in short audio clips. Additionally, the Rotary Position Embedding (RoPE)~\cite{su2024roformer} enhances the model’s ability to capture both temporal and frequency dependencies, improving its capacity to handle overlapping audio components. Fine-tuning of the model optimized its mask estimation using Time-domain MAE and Multi-resolution Complex Spectrogram MAE loss functions~\cite{guso2022loss}, which help preserve both temporal and spectral integrity. These refinements enable BS-RoFormer to effectively isolate OST from BGM while maintaining high audio quality, making it particularly suited to the complex challenges posed by short video platforms.

\subsection{Video-Music Matching}
The final module of the pipeline focuses on recommending music that aligns with video content. While the original UT-CMVMR model~\cite{unified} employed both Cross-Entropy and Triplet Loss for training, we adapted the model for short video platforms by fine-tuning it using only Triplet Loss. This modification reflects the exclusion of tag data and optimizes the similarity between video and music embeddings—treating videos as anchors, paired OSTs as positives, and non-paired BGM as negatives to enhance recommendation accuracy.

The model extracts video and audio embeddings through independent Conformer-based modules that capture both temporal and spatial dynamics. Pre-computed Optical Flow and Rhythm Quantization features, which represent motion and rhythmic characteristics, are embedded and concatenated with modality-specific embeddings. These are then normalized through fully connected layers. A Cross-Attention mechanism integrates these embeddings, learning alignment weights between modalities to capture nuanced relationships, such as rhythmic synchronization and mood alignment.

During training, Triplet Loss selects positive and negative samples based on pre-computed distances. A margin-based objective ensures that paired embeddings are closer in the latent space, while non-paired embeddings are pushed farther apart, optimizing the alignment between video and music. For this module, we use the two outputs from the mixed-music separators—arbitrary BGM and OST—as inputs. The UT-CMVMR model is then trained to classify which of these two audio components is better suited for the video content.

Figure~\ref{fig:cross_modal_pipeline} illustrates the architecture and modifications of this module. By leveraging its architectural strengths and domain-specific fine-tuning, the model effectively addresses the dynamic nature of short video platforms, ensuring accurate music recommendations.

\subsection{Integration of Modules}
The four modules are systematically integrated into a cohesive pipeline, with the output of each module serving as an essential input for the next. This modular design ensures smooth data flow and effective collaboration between components. The \textbf{Background Music Separator} and \textbf{Speech Extractor} modules work in tandem to produce clean and well-separated audio inputs, isolating vocal elements and generating `Mixed-Music' tracks. These outputs form the foundation for the \textbf{Mixed-Music Separator}, which further separates the OST and BGM components, preparing the audio data for the subsequent Video-Music Matching process.

The final module, \textbf{Video-Music Matching}, uses the refined outputs from previous stages to recommend music by aligning its rhythmic characteristics with the visual context of the video. This module is fine-tuned specifically for short video platforms, ensuring the recommendations are contextually appropriate. By integrating these modules, the pipeline can address the challenges of short video platforms, providing a robust solution for arbitrary BGM removal and OST restoration. The complete pipeline structure is illustrated in Figure~\ref{fig:end_to_end_pipeline}.

\section{Experimental Setup}
The training process for the two learned modules, Mixed-Music Separator and Video-Music Matching, took approximately 30 hours and 2 hours, respectively, utilizing a GPU cluster with NVIDIA Quadro RTX 6000 GPUs (24GB memory per GPU). The training was conducted in a Python 3.11 environment, with PyTorch 2.5.1+cu121.

For inference, processing one video through the entire pipeline takes approximately 9.17 seconds. This evaluation was performed using our curated OSVAR-160 dataset to ensure that the pipeline's performance is both realistic and comprehensive. A detailed breakdown of inference times for each module is provided in Table~\ref{tab:inference_time}.

\begin{table}[t]
\centering
\begin{tabular}{lr}
\hline
Module                  & Inference Time (s) \\ \hline
Background Music Separator       & 7.32                                 \\ 
Speech Extractor                 & 0.85                                 \\ 
Mixed-Music Separator            & 0.97                                 \\ 
Video-Music Matching             & 0.03                                 \\ \hline
\textbf{Total Pipeline Inference} & \textbf{9.17}                        \\ \hline
\end{tabular}
\caption{Inference time for each module and the total pipeline}
\label{tab:inference_time}
\vspace{-3pt}
\end{table}

\subsection{Evaluation Metrics}
This study evaluates the performance of our pipeline using metrics tailored to each module and the overall task. 

\subsubsection{Mixed-Music Separator}

The Mixed-Music Separator module is evaluated using two key metrics: Signal-to-Distortion Ratio (SDR) and Scale-Invariant SDR (SI-SDR), which measure the similarity between the reference signal and the separated signal in dB. SDR quantifies the overall quality of the separation by comparing the energy of the reference signal \(S_t\)  to that of the error signal \(S_e = S_t - S_{\text{est}}\):
\begin{equation}
\text{SDR} = 10 \cdot \log_{10} \left( \frac{\|S_t\|^2}{\|S_e\|^2} \right).
\end{equation}

Higher SDR values indicate better separation quality and signal quality. SI-SDR refines SDR by eliminating the influence of scale differences in signal amplitude, focusing solely on structural similarity. SI-SDR is computed as:
\begin{equation}
\text{SI-SDR} = 10 \cdot \log_{10} \left( \frac{\| \alpha \cdot S_t \|^2}{\| \alpha \cdot S_e \|^2} \right)
\end{equation}

where \(\alpha\) = \((\frac{S_t^\top S_{\text{est}}}{\|S_t\|^2})\). The normalization ensures robustness in scenarios with dynamic volume variations, such as those commonly encountered on short video platforms.

The two metrics complement each other; SDR evaluates overall quality, while SI-SDR isolates structural differences, providing a robust and comprehensive evaluation of the separator’s performance in restoring high-quality audio under diverse conditions.

\subsubsection{Video-Music Matching}
The Video-Music Matching module is evaluated using accuracy, which measures the proportion of correctly identified matches between videos and their intended music (i.e., OST or mismatched BGM). This task is framed as a binary classification problem, where the model determines whether the video is paired with the appropriate music, either the OST or the arbitrary BGM. While some prior studies employed other metrics for retrieval performance, accuracy was selected for this study because it directly aligns to validate precise alignment between video content and music recommendations. This metric ensures that the model not only identifies the correct music but also does so without relying on ranking, which is crucial for real-time applications on short video platforms.


\subsubsection{Pipeline Evaluation: Arbitrary BGM Removal}
To evaluate the pipeline’s overall performance, particularly for arbitrary BGM removal, the same metrics used for the Mixed-Music Separator are applied (SDR and SI-SDR). These metrics assess the system's ability to separate audio components while preserving their temporal and spectral integrity. The use of SDR and SI-SDR ensures that both separation quality and restoration fidelity are rigorously validated.

\section{Results}
\subsection{Mixed-Music Separator}
The evaluation of the Mixed-Music Separator module shows significant improvements in OST and BGM separation, owing to domain-specific fine-tuning on the OASD-20K dataset. The proposed approach achieves an average SDR of 11.45 dB and SI-SDR of 11.06 dB for BGM separation, and an SDR of 8.23 dB and SI-SDR of 7.53 dB for OST separation.

For comparison, the original BS-RoFormer model, trained on a different dataset with 6-second audio clips, achieves an average SDR of 9.76 dB across multi-component separation tasks (e.g., vocals, drums, bass). While these tasks and datasets differ, the improved performance highlights the effectiveness of domain-specific fine-tuning for OST-BGM separation in the context of short video platforms.

As shown in Table~\ref{tab:compact_comparison}, the fine-tuned model demonstrates robust performance in addressing complex overlaps and transitions characteristic of short-form content.

\begin{table}[t]
\centering
\begin{tabular}{lrr}
\hline
Metric           & Fine-Tuned (dB) \(\uparrow\) & Non-Tuned (dB) \(\uparrow\)  \\ \hline
BGM SDR          & \textbf{11.45}               & 2.36           \\ 
BGM SI-SDR       & \textbf{11.06}               & 2.13           \\ 
OST SDR          & \textbf{8.23}                & 0.73           \\ 
OST SI-SDR       & \textbf{7.53}                & 0.54           \\ \hline
\end{tabular}
\caption{Comparison of fine-tuned and non-tuned BS-RoFormer performance with SDR and SI-SDR metrics. \(\uparrow\) indicates that higher values are better.}
\label{tab:compact_comparison}
\end{table}

The improvements in SDR and SI-SDR values indicate that the separated audio components closely resemble the original signals, particularly for BGM separation, where the model demonstrates a strong capability to handle frequent overlaps and rapid transitions typical of short-form content. While the OST separation scores are slightly lower—likely due to the inherent complexity and variability of OST signals—the overall results affirm that BS-RoFormer, with domain-specific fine-tuning, effectively addresses the unique challenges of audio separation in short video platforms. These findings underscore the model's adaptability and robustness in real-world audio environments.

\subsection{Video-Music Matching}
The evaluation of the Video-Music Matching module demonstrates the superior performance of the UT-CMVMR model compared to the CBVMR baseline. Specifically, UT-CMVMR achieves an impressive accuracy of 96.8\%, significantly outperforming CBVMR's 65.4\%, as detailed in Table~\ref{tab:video_music_matching}. This improvement underscores the model's capability to effectively align video and audio content, even in metadata-scarce environments such as YouTube Shorts and TikTok.

\begin{table}[t]
\centering
\begin{tabular}{lr}
\hline
Model      & Accuracy (\%) \(\uparrow\) \\ \hline
CBVMR \cite{hong2018cbvmr}    & 65.4                     \\ 
UT-CMVMR \cite{unified} & \textbf{96.8} \\ \hline
\end{tabular}
\caption{Comparison of matching accuracy.}
\label{tab:video_music_matching}
\end{table}

UT-CMVMR’s strong performance can be attributed to its architectural innovations, such as the integration of Optical Flow and Rhythm Quantization modules, which enhance its ability to capture temporal dynamics and rhythmic patterns. The use of advanced cross-modal attention mechanisms further facilitates effective interaction between video and audio embeddings, enabling precise alignment. Fine-tuning on the domain-specific HIMV-200K dataset equips the model to effectively address the unique challenges of short video platforms, ensuring it can recommend rhythmically and contextually appropriate audio.

The high accuracy achieved by UT-CMVMR underscores its effectiveness in handling user-generated content and its practical utility in real-world applications, validating the benefits of domain-specific fine-tuning and architectural enhancements for short video platforms.

\subsection{Pipeline Performance}
The evaluation of arbitrary BGM separation across three distinct scenarios demonstrates that ours consistently achieves superior performance metrics compared to alternative approaches. These scenarios were designed to benchmark the pipeline's ability to reconstruct audio by comparing the original reference signals with: (1) The final output of ours, which separates arbitrary BGM while restoring OST, (2) Mixed-music, where arbitrary BGM is added to the original OST, and (3) Dialogue-only audio, where BGM is removed entirely, isolating speech. Each scenario was evaluated against the reference signals using SDR and SI-SDR, offering a direct measure of audio quality and structural similarity.

Our approach achieves an average \textbf{SDR of 138.63 dB} and \textbf{SI-SDR of 16.53 dB}, significantly outperforming the other evaluation scenarios. As shown in Table~\ref{tab:short_video_results}, when arbitrary BGM is added to the original OST, the resulting mixed audio demonstrates a high \textbf{SDR of 129.51 dB} but a considerably lower \textbf{SI-SDR of 8.65 dB}. This indicates that while the overall signal remains clear and loud, the structural quality relative to the original is compromised due to the interference process from the added BGM. The SI-SDR metric here is crucial, as it emphasizes the intrinsic similarity between the reference and estimated signals, unaffected by amplitude discrepancies.

\begin{table}[t]
\begin{tabular}{%
  >{\raggedright\arraybackslash}m{3.25cm}   
  >{\raggedleft\arraybackslash}m{1.75cm}   
  >{\raggedleft\arraybackslash}m{2.25cm}} 
\hline
Scenario & Average SDR (dB)\,$\uparrow$ & Average SI-SDR (dB)\,$\uparrow$ \\ \hline
\textbf{Ours}--Arbitrary BGM Separation & \textbf{138.63} & \textbf{16.53} \\ \hline
Audio with Arbitrary BGM Added          & 129.51          & 8.65           \\ \hline
Dialogue-Only Audio (BGM Removed)       & 131.24          & 14.00          \\ \hline
\end{tabular}
\caption{Performance comparison across different audio scenarios.}
\label{tab:short_video_results}
\vspace{-3pt}
\end{table}

When comparing arbitrary BGM separation to the dialogue-only audio scenario (which achieves an average \textbf{SDR of 131.24 dB} and \textbf{SI-SDR of 14.00 dB}), ours demonstrates slightly better performance. This showcases the model's ability not only to effectively remove the BGM but also to reconstruct the OST in a manner that closely matches the original audio, significantly preserving essential spectral and temporal details. These results validate our effectiveness in maintaining both high-quality and structural similarity across diverse audio scenarios. By comparing the restored outputs to the original reference audio, the evaluation ensures that SDR and SI-SDR metrics reliably reflect the model’s capacity to accurately reconstruct audio components.

\section{Conclusion}
This study proposes a novel pipeline designed to address the challenge of removing arbitrarily inserted BGM in short video platforms. By integrating advanced models with domain-specific fine-tuning, the pipeline demonstrates robust performance in separating OSTs from BGM while effectively restoring the original audio. The Mixed-Music Separator, fine-tuned on the domain-specific \textbf{OASD-20K} dataset, efficiently handles complex audio blends, achieving high SDR and SI-SDR values. The pipeline's final module, Video-Music Matching, ensures precise alignment between audio and video content, leveraging the comprehensive HIMV-200K dataset for fine-tuning. Additionally, the \textbf{OSVAR-160} dataset serves as a benchmark for evaluating the pipeline under realistic short video scenarios.

In addition, we developed two novel datasets: \textbf{OASD-20K}, comprising 20,000 arbitrary BGM and OST mixed audio clips, and \textbf{OSVAR-160}, consisting of 1,121 video and mixed-audio pairs. These domain-specific datasets serve both as a resource for mixed-music separation tasks and as a benchmark for evaluating audio restoration in short video content.

While our approach achieves strong results, there remain opportunities for improvement. Future work could focus on enhancing the preservation of non-musical audio effects and further improving the quality of separated audio. Expanding the dataset to include diverse genres and languages will also enhance the adaptability across various global platforms.

\clearpage

\section*{Acknowledgements}

This research was supported by the MSIT (Ministry of Science and ICT), Korea, under the ICAN (ICT Challenge and Advanced Network of HRD) support program (RS-2024-00436934), and the ITRC (Information Technology Research Center) program (IITP-2025-RS-2024-00436936) supervised by the IITP (Institute for Information \& Communications Technology Planning \& Evaluation). Moreover, this work was supported by the IITP grant funded by the MSIT (RS-2024-00439139, Development of a Cyber Crisis Response and Resilience Test Evaluation Systems).

\bibliographystyle{named}
\bibliography{ijcai25}

\begin{thebibliography}{}

\bibitem[\protect\citeauthoryear{AL and Chopra}{2024}]{al2024regulating}
Reeta~Sony AL and Shruti Chopra.
\newblock Regulating digital era: A comparative analysis of policy perspectives on media entertainment.
\newblock {\em Legal Issues Digit. Age}, 5:97, 2024.

\bibitem[\protect\citeauthoryear{Avgoustinakis \bgroup \em et al.\egroup }{2021}]{avgoustinakis2021audio}
Pavlos Avgoustinakis, Giorgos Kordopatis-Zilos, Symeon Papadopoulos, Andreas~L Symeonidis, and Ioannis Kompatsiaris.
\newblock Audio-based near-duplicate video retrieval with audio similarity learning.
\newblock In {\em 2020 25th International Conference on Pattern Recognition (ICPR)}, pages 5828--5835. IEEE, 2021.

\bibitem[\protect\citeauthoryear{Blanding}{2024}]{blanding2024navigating}
Erin Blanding.
\newblock Navigating the social media landscape.
\newblock \url{https://dukespace.lib.duke.edu/server/api/core/bitstreams/72552abe-d01a-4701-9ef3-429cef476b8f/content}, 2024.

\bibitem[\protect\citeauthoryear{Bredin \bgroup \em et al.\egroup }{2020}]{bredin2020pyannote}
Herv{\'e} Bredin, Ruiqing Yin, Juan~Manuel Coria, Gregory Gelly, Pavel Korshunov, Marvin Lavechin, Diego Fustes, Hadrien Titeux, Wassim Bouaziz, and Marie-Philippe Gill.
\newblock Pyannote. audio: neural building blocks for speaker diarization.
\newblock In {\em ICASSP 2020-2020 IEEE International Conference on Acoustics, Speech and Signal Processing (ICASSP)}, pages 7124--7128. IEEE, 2020.

\bibitem[\protect\citeauthoryear{Cano \bgroup \em et al.\egroup }{2018}]{cano2018musical}
Estefania Cano, Derry FitzGerald, Antoine Liutkus, Mark~D Plumbley, and Fabian-Robert St{\"o}ter.
\newblock Musical source separation: An introduction.
\newblock {\em IEEE Signal Processing Magazine}, 36(1):31--40, 2018.

\bibitem[\protect\citeauthoryear{Charfeddine \bgroup \em et al.\egroup }{2022}]{audiowatermark}
Maha Charfeddine, Eya Mezghani, Salma Masmoudi, Chokri~Ben Amar, and Hesham Alhumyani.
\newblock Audio watermarking for security and non-security applications.
\newblock {\em IEEE Access}, 10:12654--12677, 2022.

\bibitem[\protect\citeauthoryear{Cho \bgroup \em et al.\egroup }{2014}]{cho2014learning}
Kyunghyun Cho, Bart van Merrienboer, Caglar Gulcehre, Dzmitry Bahdanau, Fethi Bougares, Holger Schwenk, and Yoshua Bengio.
\newblock Learning phrase representations using rnn encoder--decoder for statistical machine translation.
\newblock In {\em Proceedings of the 2014 Conference on Empirical Methods in Natural Language Processing (EMNLP)}, pages 1724--1734, 2014.

\bibitem[\protect\citeauthoryear{Cohen and Berdugo}{2001}]{cohen2001speech}
Israel Cohen and Baruch Berdugo.
\newblock Speech enhancement for non-stationary noise environments.
\newblock {\em Signal processing}, 81(11):2403--2418, 2001.

\bibitem[\protect\citeauthoryear{D{\'e}fossez \bgroup \em et al.\egroup }{2019}]{demucs}
Alexandre D{\'e}fossez, Nicolas Usunier, L{\'e}on Bottou, and Francis Bach.
\newblock Demucs: Deep extractor for music sources with extra unlabeled data remixed.
\newblock \url{https://arxiv.org/abs/1909.01174}, 2019.

\bibitem[\protect\citeauthoryear{{DemandSage}}{2023}]{demandsage:2023}
{DemandSage}.
\newblock Video marketing statistics: Trends and insights for 2023, 2023.
\newblock Accessed: 2023-12-31.

\bibitem[\protect\citeauthoryear{EBU-Recommendation}{2011}]{ebu2011loudness}
R~EBU-Recommendation.
\newblock Loudness normalisation and permitted maximum level of audio signals.
\newblock \url{https://www.rosseladvertising.be/sites/default/files/techspecs/TV/ebu_recommendation.pdf}, 2011.

\bibitem[\protect\citeauthoryear{Gu \bgroup \em et al.\egroup }{2023}]{gu2023dual}
Xin Gu, Yinghua Shen, and Chaohui Lv.
\newblock A dual-path cross-modal network for video-music retrieval.
\newblock {\em Sensors}, 23(2):805, 2023.

\bibitem[\protect\citeauthoryear{Gus{\'o} \bgroup \em et al.\egroup }{2022}]{guso2022loss}
Enric Gus{\'o}, Jordi Pons, Santiago Pascual, and Joan Serr{\`a}.
\newblock On loss functions and evaluation metrics for music source separation.
\newblock In {\em ICASSP 2022-2022 IEEE International Conference on Acoustics, Speech and Signal Processing (ICASSP)}, pages 306--310. IEEE, 2022.

\bibitem[\protect\citeauthoryear{Han \bgroup \em et al.\egroup }{2024}]{han2024sfd}
Chaowei Han, Gaofeng Meng, and Chunlei Huo.
\newblock Sfd: Similar frame dataset for content-based video retrieval.
\newblock In {\em 2024 IEEE International Conference on Image Processing (ICIP)}, pages 2403--2409. IEEE, 2024.

\bibitem[\protect\citeauthoryear{Hasanovic \bgroup \em et al.\egroup }{2022}]{hasanovic2022can}
Adi Hasanovic, P~Halliday, and MHM Schellekens.
\newblock How can online creators with derivative creative works on youtube protect themselves from questionable usage of intellectual property rights?
\newblock \url{https://arno.uvt.nl/show.cgi?fid=157488}, 2022.

\bibitem[\protect\citeauthoryear{Hennequin \bgroup \em et al.\egroup }{2020}]{spleeter}
Romain Hennequin, Anis Khlif, Felix Voituret, and Manuel Moussallam.
\newblock Spleeter: a fast and efficient music source separation tool with pre-trained models.
\newblock {\em Journal of Open Source Software}, 5(50):2154, 2020.

\bibitem[\protect\citeauthoryear{Henry \bgroup \em et al.\egroup }{2024}]{henry2024fast}
Chris Henry, Li~Song, and Zhu Li.
\newblock Fast video deduplication and localization with temporal consistence re-ranking.
\newblock {\em IEEE Transactions on Circuits and Systems for Video Technology}, 34(11):12006--12018, 2024.

\bibitem[\protect\citeauthoryear{Hong \bgroup \em et al.\egroup }{2018}]{hong2018cbvmr}
Sungeun Hong, Woobin Im, and Hyun~S Yang.
\newblock Cbvmr: content-based video-music retrieval using soft intra-modal structure constraint.
\newblock In {\em Proceedings of the 2018 ACM on international conference on multimedia retrieval}, pages 353--361, 2018.

\bibitem[\protect\citeauthoryear{Huang \bgroup \em et al.\egroup }{2014}]{huang2014deep}
Po-Sen Huang, Minje Kim, Mark Hasegawa-Johnson, and Paris Smaragdis.
\newblock Deep learning for monaural speech separation.
\newblock In {\em 2014 IEEE International Conference on Acoustics, Speech and Signal Processing (ICASSP)}, pages 1562--1566. IEEE, 2014.

\bibitem[\protect\citeauthoryear{Jansson \bgroup \em et al.\egroup }{2017}]{jansson2017singing}
A~Jansson, E~Humphrey, N~Montecchio, R~Bittner, A~Kumar, and T~Weyde.
\newblock Singing voice separation with deep u-net convolutional networks.
\newblock In {\em 18th International Society for Music Information Retrieval Conference}, pages 23--27, 2017.

\bibitem[\protect\citeauthoryear{Jin \bgroup \em et al.\egroup }{2021}]{jin2021cross}
Cong Jin, Tian Zhang, Shouxun Liu, Yun Tie, Xin Lv, Jianguang Li, Wencai Yan, Ming Yan, Qian Xu, Yicong Guan, et~al.
\newblock Cross-modal deep learning applications: audio-visual retrieval.
\newblock In {\em Pattern Recognition. ICPR International Workshops and Challenges: Virtual Event, January 10--15, 2021, Proceedings, Part VI}, pages 301--313. Springer, 2021.

\bibitem[\protect\citeauthoryear{Kim and Reiter}{2017}]{soo2017interpretable}
Tae~Soo Kim and Austin Reiter.
\newblock Interpretable 3d human action analysis with temporal convolutional networks.
\newblock In {\em Proceedings of the IEEE conference on computer vision and pattern recognition workshops}, pages 20--28, 2017.

\bibitem[\protect\citeauthoryear{Li and Kumar}{2019}]{li2019query}
Bochen Li and Aparna Kumar.
\newblock Query by video: Cross-modal music retrieval.
\newblock In {\em 20th International Society for Music Information Retrieval Conference}, pages 604--611, 2019.

\bibitem[\protect\citeauthoryear{Liu \bgroup \em et al.\egroup }{2010}]{liu2010coherent}
Yang Liu, Wan-Lei Zhao, Chong-Wah Ngo, Chang-Sheng Xu, and Han-Qing Lu.
\newblock Coherent bag-of audio words model for efficient large-scale video copy detection.
\newblock In {\em Proceedings of the ACM international conference on image and video retrieval}, pages 89--96, 2010.

\bibitem[\protect\citeauthoryear{Lu \bgroup \em et al.\egroup }{2024}]{lu2024music}
Wei-Tsung Lu, Ju-Chiang Wang, Qiuqiang Kong, and Yun-Ning Hung.
\newblock Music source separation with band-split rope transformer.
\newblock In {\em ICASSP 2024-2024 IEEE International Conference on Acoustics, Speech and Signal Processing (ICASSP)}, pages 481--485. IEEE, 2024.

\bibitem[\protect\citeauthoryear{Luo and Yu}{2023}]{luo2023music}
Yi~Luo and Jianwei Yu.
\newblock Music source separation with band-split rnn.
\newblock {\em IEEE/ACM Transactions on Audio, Speech, and Language Processing}, 31:1893--1901, 2023.

\bibitem[\protect\citeauthoryear{Mao \bgroup \em et al.\egroup }{2024}]{unified}
Tianjun Mao, Shansong Liu, Yunxuan Zhang, Dian Li, and Ying Shan.
\newblock Unified pretraining target based video-music retrieval with music rhythm and video optical flow information.
\newblock In {\em ICASSP 2024-2024 IEEE International Conference on Acoustics, Speech and Signal Processing (ICASSP)}, pages 7890--7894. IEEE, 2024.

\bibitem[\protect\citeauthoryear{{Market Research Future}}{2023}]{marketresearchfuture:2023}
{Market Research Future}.
\newblock Short video platform market size, share, and report 2032, 2023.
\newblock Accessed: 2023-12-31.

\bibitem[\protect\citeauthoryear{Pr{\'e}tet \bgroup \em et al.\egroup }{2021}]{pretet2021cross}
Laure Pr{\'e}tet, Gael Richard, and Geoffroy Peeters.
\newblock Cross-modal music-video recommendation: A study of design choices.
\newblock In {\em 2021 International Joint Conference on Neural Networks (IJCNN)}, pages 1--9. IEEE, 2021.

\bibitem[\protect\citeauthoryear{Ronneberger \bgroup \em et al.\egroup }{2015}]{ronneberger2015u}
Olaf Ronneberger, Philipp Fischer, and Thomas Brox.
\newblock U-net: Convolutional networks for biomedical image segmentation.
\newblock In {\em Medical image computing and computer-assisted intervention--MICCAI 2015: 18th international conference, Munich, Germany, October 5-9, 2015, proceedings, part III 18}, pages 234--241. Springer, 2015.

\bibitem[\protect\citeauthoryear{Saracoglu \bgroup \em et al.\egroup }{2009}]{saracoglu2009content}
Ahmet Saracoglu, Ersin Esen, Tugrul~K Ates, Banu~Oskay Acar, Unal Zubari, Ezgi~C Ozan, Egemen Ozalp, A~Aydin Alatan, and Tolga Ciloglu.
\newblock Content based copy detection with coarse audio-visual fingerprints.
\newblock In {\em 2009 Seventh International Workshop on Content-Based Multimedia Indexing}, pages 213--218. IEEE, 2009.

\bibitem[\protect\citeauthoryear{Sharif~Razavian \bgroup \em et al.\egroup }{2014}]{sharif2014cnn}
Ali Sharif~Razavian, Hossein Azizpour, Josephine Sullivan, and Stefan Carlsson.
\newblock Cnn features off-the-shelf: an astounding baseline for recognition.
\newblock In {\em Proceedings of the IEEE conference on computer vision and pattern recognition workshops}, pages 806--813, 2014.

\bibitem[\protect\citeauthoryear{Su \bgroup \em et al.\egroup }{2024}]{su2024roformer}
Jianlin Su, Murtadha Ahmed, Yu~Lu, Shengfeng Pan, Wen Bo, and Yunfeng Liu.
\newblock Roformer: Enhanced transformer with rotary position embedding.
\newblock {\em Neurocomputing}, 568:127063, 2024.

\bibitem[\protect\citeauthoryear{Udaypur}{2024}]{udaypur2024novel}
Bangalore Udaypur.
\newblock A novel dc-gcn with attention mechanism for accurate near-duplicate video data cleaning.
\newblock \url{https://anapub.co.ke/journals/jmc/jmc_pdf/2024/jmc_volume_4-issue_4/JMC202404093.pdf}, 2024.

\bibitem[\protect\citeauthoryear{Virtanen}{2007}]{virtanen2007monaural}
Tuomas Virtanen.
\newblock Monaural sound source separation by nonnegative matrix factorization with temporal continuity and sparseness criteria.
\newblock {\em IEEE transactions on audio, speech, and language processing}, 15(3):1066--1074, 2007.

\bibitem[\protect\citeauthoryear{Won \bgroup \em et al.\egroup }{2021}]{won2021multimodal}
Minz Won, Sergio Oramas, Oriol Nieto, Fabien Gouyon, and Xavier Serra.
\newblock Multimodal metric learning for tag-based music retrieval.
\newblock In {\em ICASSP 2021-2021 IEEE International Conference on Acoustics, Speech and Signal Processing (ICASSP)}, pages 591--595. IEEE, 2021.

\bibitem[\protect\citeauthoryear{Yi \bgroup \em et al.\egroup }{2021}]{yi2021cross}
Jing Yi, Yaochen Zhu, Jiayi Xie, and Zhenzhong Chen.
\newblock Cross-modal variational auto-encoder for content-based micro-video background music recommendation.
\newblock {\em IEEE Transactions on Multimedia}, 25:515--528, 2021.

\bibitem[\protect\citeauthoryear{{YouTube Help}}{2023}]{youtubehelp:shorts}
{YouTube Help}.
\newblock Music eligibility for youtube shorts, 2023.
\newblock Accessed: 2023-12-31.

\bibitem[\protect\citeauthoryear{Zhao and Guo}{2024}]{zhao2024liability}
Yun Zhao and Yijin Guo.
\newblock Liability regulation on short video platforms: balancing freedom of expression and copyright protection.
\newblock {\em International Journal of Legal Discourse}, 9(2):313--338, 2024.

\end{thebibliography}

\end{document}